\documentclass[final,3p,times,twocolumn]{elsarticle}

\usepackage{graphicx}

\usepackage{amsmath,amssymb}
\journal{Physica E}

\newcommand{\mr}{\mathrm}

\begin{document}

\begin{frontmatter}



\title{Manifestation of anomalous Floquet states with longevity in dynamic fractional Stark ladder with high AC electric fields}


\author[1]{Yuya Nemoto}      
\author[1]{Fumitaka Ohno}
\author[2,3]{Nobuya Maeshima\corref{cor}}
\ead{maeshima@ims.tsukuba.ac.jp}
\author[2,3]{Ken-ichi Hino}
\cortext[cor]{Corresponding author}

\address[1]{Doctoral Program in Materials Science, Graduate School of Pure and Applied Sciences, University of Tsukuba, Tsukuba, Ibaraki 305-8573, Japan}
\address[2]{Division of Materials Science, Faculty of Pure and Applied Sciences, University of Tsukuba, Tsukuba 305-8573, Japan}
\address[3]{Center for Computational Sciences, University of Tsukuba, Tsukuba 305-8577, Japan}

\begin{abstract}
We examine a resonance structure of Floquet state in dynamic fractional Stark ladder (DFSL) realized in biased semiconductor superlattices driven by a terahertz cw laser on the basis of the R-matrix Floquet theory. 
To do this, we calculate an excess density of state $\rho^{(ex)}(E)$ corresponding to lifetime of the Floquet state with a fractional matching ratio $\eta$, where $\eta$ is the ratio of a Bloch frequency $\Omega_{B}$ to a laser frequency $\omega$, namely, $\eta=\Omega_{B}/\omega$. 
The results for $\eta=3/2$ demonstrate the appearance of discernibly large peaks associated with Floquet states with longevity in a region of relatively high laser-intensity.
The underlying physics is discussed in terms of an analytical expression of $\rho^{(ex)}(E)$ and the associated Green function in which ponderomotive couplings are included in a non-perturbative way.

\end{abstract}


\end{frontmatter}


\section{Introduction}
\label{sect1}

Control of quantum states by AC electric fields has attracted considerable attention in condensed matter physics~\cite{kohler,grifoni,yonemitsu1}.
In particular, coherent control of single-electron states in semiconductor superlattices (SLs) is a longstanding issue.  It can be easily dealt with as a single-electron problem in a periodic potential driven by external fields, and has provided fascinating phenomena.
A typical example of the related phenomena is the dynamical Wannier-Stark ladder (DWSL) in biased semiconductor SLs driven by a terahertz cw laser\cite{holthaus,zak,yashima,zhao,je,gluck}. 
As is well known, when a static electric field $F_0$ is applied to semiconductor SLs with a lattice constant $d$, there appears discretized energy levels 
(Wannier-Stark ladder, abbreviated as WSL) with a constant energy interval of a Bloch frequency $\Omega_B=edF_0/\hbar$, 
where $e$ and $\hbar$ represent the elementary  charge of electron and Planck's constant divided by $2\pi$, respectively.
DWSL is the WSL driven by a cw laser with the frequency $\omega$.
The applied laser generally causes the photon-assisted tunneling (PAT) between discrete WSL levels, leading to miniband formation accompanied by delocalized states of electron.
On the contrary, at a certain strength of the laser intensity the formed miniband is collapsed, resulting in dynamic localization of electrons.
Thus, in DWSL, both of the localized and delocalized states are coherently controlled by altering a couple of laser parameters.

As stated above, DWSL can be treated as a single-electron problem and has often been investigated by using a tight binding model.
Here, it should be noted that the tight binding model only takes account of a few lowest bound states in quantum wells.
In fact, in a real situation, a strong laser field induces non-negligible couplings not only between these bound states but among all states including all bound states and continuum states.  The effects caused by these couplings  are termed as ac-Zener tunneling (ac-ZT) effects, and the tight binding model only partially incorporates the ac-ZT effects~\cite{rotvig,holthaus2}.
Some of the present authors and coworkers treat DWSL as a multichannel scattering problem incorporating all parts of the ac-ZT effects~\cite{kukuu,nemoto},
and investigate the resonances structure of DWSL with integer numbers of a ratio $\eta=\Omega_B/\omega$ by virtue of the numerical method based on the R-matrix Floquet theory~\cite{RFT}; $\eta$ is termed the matching ratio.
Here the term "channel" is associated with a photon-sideband state formed in DWSL.
It is well-known that physical properties of DWSL are characterized by the ratio $\eta$.

DWSL with fractional $\eta=p/q$ is an objective of the present study, where $p$ and $q$ are relatively prime numbers.
Herein, this is called dynamic fractional Stark ladder (DFSL)~\cite{zhao}.
In particular, we focus on multi-channel resonance structure of DFSL caused by the ac-ZT effect in a region of high laser-intensity.
According to the preceding study~\cite{karasawa}, it is shown that a resonance peak of DFSL is shifted and broadened by the effect of the ac-ZT and a fractal pattern characteristic of DFSL~\cite{zhao} does not necessarily hold. 
However, this study resorts just to a single-channel scattering problem without interchannel couplings; 
this is considered correct just under the high-frequency approximation with $\omega\to\infty$.

We examine the resonance structure of DFSL by resorting to a multichannel scattering problem including full effects of interchannel couplings.
Here,  excess density of states (DOS)~\cite{exdos1,exdos2,exdos3}, $\rho^{(ex)}(E)$, is considered, which is closely related to the lifetime of the concerned Floquet state with quasienergy $E$.
The obtained results are discussed with the idea that $\rho^{(ex)}(E)$ is analytically classified into a single-channel contribution $\rho^{(ex)}_0(E)$, a multichannel non-resonance term $\rho^{(ex)}_{nr}(E)$, and a multichannel resonance term $\rho^{(ex)}_{res}(E)$~\cite{nemoto}. 
For $\eta=3/2$, $\rho^{(ex)}(E)$ shows the appearance of a larger number of new peaks with increasing laser-intensity. 
In particular, we put a stress on that one of these grows into a new and discernibly intense peak in the vicinity of the second-lowest level of the WSL with anomalously long lifetime. 
%For $\eta=3/2$, $\rho^{(ex)}(E)$ shows anomalous behavior 
As shown later, these phenomena are explained based on the mechanism of Feshbach-like resonance~\cite{Feshbach}. 

%, and are not observed in DWSL with integer $\eta$~\cite{nemoto,kukuu}.
%here �� is the ratio of a Bloch frequency ��B to a laser frequency ��, namely, ��=��_B/��. When �� is equal to a positive integer (this type of DWSL is termed the integer type DWSL in this study), photon assisted tunneling (PAT) and DL are realized. PAT is a tunneling   For fractional ��=p/q with p and q as prime numbers (this type of DWSL is termed the fractional type DWSL in this study), it is reported that the fractal structure exists in a E versus ��^(-1) diagram with E as quasienergy [9]. 
%Physical properties of the DWSL are known to be governed by the matching ratio 
%Some of the present authors have investigated resonance structure of DWSL with integer $\eta$ by using a numerical technique based on the multichannel scattering theory, and found that ac-Zener tunneling (ac-ZT) causes destabilization and deformation of the resonance structure in accordance with the mechanism of Feshbach resonance in high-intensity laser region~\cite{nemoto}. 

\section{Theory}
\label{sect2}

The theoretical framework of our study is already described in the series of preceding papers~\cite{kukuu,nemoto}.
In this section, we present its brief explanation for readers' understanding.

\label{sec2A}
We start with the DWSL Hamiltonian,
\begin{equation}
H(z,t)=\left[ p_z + \frac{1}{c} A(t) \right]\frac{1}{2m(z)}\left[ p_z + \frac{1}{c} A(t) \right] + V(z),
\label{Hamiltonian}
\end{equation}
where $V(z)$, $m(z)$ and $p_z$ represent a confining potential of a semiconductor SLs, an effective mass of electron, and a momentum operator along the crystal growth direction $z$, respectively. In addition, $A(t)$ is a vector potential at time $t$ for an applied electric field $F(t)=-\dot{A}(t)/c$, where $c$ is the speed of light. In the following, we use the atomic unit.
By applying a unitary transformation called the Kramers-Henneberger transformation~\cite{gavrila,KH}, the Hamiltonian of Eq. (\ref{Hamiltonian}) is rewritten as
\begin{align}
\mathcal{H}(z,t)=p_z \left\{ {1 \over 2m[z+a(t)]} \right\} p_z +V[z+a(t)] \nonumber\\
 +F_0z+v(z,t), 
\label{barHx}
\end{align}
where $v(z,t)$ is a residual interaction caused by the $z$ dependence of the effective mass $m(z)$~\cite{kukuu}.
Here, $a(t) \equiv \alpha \mr{cos}\omega t$ is the position of a classical electron oscillating under a laser field 
$F(t)=F_{ac} \mr{cos}{\omega t}$, and $\alpha$ defined as
\begin{equation}
\alpha=\frac{F_{ac}}{m_{as} \omega^2}
\end{equation}
is a ponderomotive radius corresponding to the excursion amplitude of a classical electron.
It is remarked that, in the asymptotic region of $|z|>>L_{SL}$, $V[z+a(t)]$ and $m[z+a(t)]$ become the 
constant values of $V_{as}$ and $m_{as}$, respectively. Thus, $v(z,t)$ vanishes in the asymptotic region.
Here $L_{SL}$ represents the size of SL.

The Hamiltonian $\mathcal{H}(z,t)$ satisfies the Schr\"{o}dinger equation
\begin{equation}
\left[\mathcal{H}(z,t)-i{\partial \over \partial t}\right]\Phi(z,t)=0,
\label{barH}
\end{equation}
and ensures the Floquet theorem (see \ref{sec_app_A}) because of the temporal periodicity $\mathcal{H}(z,t+T)=\mathcal{H}(z,t)$ with the period $T=2\pi/\omega$.
Then, the wave function $\Phi(z,t)$ is expressed as
\begin{equation}
\Phi(z,t)=\exp{(-iEt)} \sum_{\nu =- N_{ph} }^{N_{ph}} \exp{(i\nu \omega t)}\psi_\nu(z),
\label{eq_fourier}
\end{equation}
and  Eq. (\ref{barH}) is reduced into coupled differential equations for a set of wavefunction $\{ \psi_\nu\}$
\begin{equation}
\sum_{ \nu =-N_{ph} }^{N_{ph}} \left[ L_{\mu \nu}(z)-E\delta_{\mu \nu}\right]\psi_{\nu}(z)=0,
\label{L}
\end{equation}
where $N_{ph}$ represents a cut-off of the photon index $\nu$ with $N_{ph}>>1$, and $L_{\mu \nu}(z)$ is given by $L_{\mu \nu}(z)=\mathcal{H}_{\mu \nu}(z) + \nu \omega\delta_{\mu \nu}$.
Here, the matrix element $O_{\mu\nu}(z)$ is defined as
\begin{equation}
O_{\mu \nu}(z)={1 \over T} \int^{T}_0 \exp{[-i(\mu -\nu)\omega t]} O(z,t) dt.
\end{equation}
It is noted that $O_{\mu \nu}(z)$ is a function of $\alpha$ as well as $z$, though the notation of $\alpha$ is omitted for the sake of simplicity.

In our framework, Eq. (\ref{L}) can be dealt with as a multichannel scattering problem, where there are $N_{ch}$ independent solutions $\{ \psi_{\nu \beta}(z)\}$ with $\nu$, $\beta = 1 \sim N_{ch}$; $N_{ch}$ is the number of channels defined as $N_{ch}=2N_{ph}+1$~\cite{ochan}.
To be concrete, $\psi_{\nu \beta}(z)$ represents the $\nu$-th channel of the $\beta$-th solution.
In the asymptotic region of $|z|>>L_{SL}$, the Hamiltonian of Eq. (\ref{barHx}) is reduced to
\begin{align}
\mathcal{H}(z,t) \to \mathcal{H}_{as}(z)=\frac{p_z^2}{2m_{as}} +F_0z + V_{as}
\label{asym}
\end{align}
and thus, the interchannel component of $L_{\mu \nu}(z)$ diminishes and $L_{\mu \nu}(z)$ becomes
\begin{equation}
  L_{\mu \nu}(z) \to \left[ \mathcal{H}_{as}(z) + \nu \omega \right]\delta_{\mu \nu}.
\end{equation}
In the region of $|z|>>L_{SL}$ with negative $z$, the scattering boundary conditions
\begin{equation}
\psi_{\nu \beta}(z)=\chi^{(+)}_{\nu\beta}(z)-\sum_{\beta^\prime=1}^{N_{ch}}\chi^{(-)}_{\nu\beta^\prime}(z) S_{\beta^\prime \beta}(E)
\label{boundary1}
\end{equation}
is imposed on $\psi_{\nu \beta}$, where $S(E)$ represents a scattering matrix, $\chi^{(\pm)}_{\nu\beta}(z)=\phi^{(\pm)}_\nu(z)\delta_{\nu\beta}$, and 
$\phi^{(\pm)}_\nu(z)$ is an energy-normalized progressive wave in the direction of $\pm z$,
satisfying the asymptotic field-free equation 
\begin{equation}
\left[   \mathcal{H}_{as}(z) + \nu \omega -  E \right] \phi^{(\pm)}_\nu(z) = 0
  \label{eq_aseq}
\end{equation}
associated with an Airy function. In addition, $\psi_{\nu \beta}(z)$ should vanish in the region of $z >> L_{SL}$.

The scattering matrix $S(E)$ is calculated numerically by employing the R-matrix propagation techniques~\cite{Rpro}, 
and the key value of excess DOS is obtained by 
\begin{equation}
\rho^{(ex)} (E)={i\over 2} {\rm Tr} \left\{[S(E)]^{-1} { d S(E) \over d E}\right\}.
\label{rho}
\end{equation}
Here we note that this is also expressed as
\begin{equation}
  \rho^{(ex)}(E)  = \rho (E) - \rho^{(as)}(E),
\end{equation}
where $\rho(E)$ represents the DOS of the DWSL, and $\rho^{(as)}$ is the DOS of a field-free asymptotic state $\chi^{(\pm)}_{\nu\beta}(z)$.
In addition, $\rho^{(ex)} (E)$ is related to the lifetime of the concerned Floquet state with $E$~\cite{exdos1,exdos2,exdos3};
\begin{equation}
\tau(E)  = \frac{\rho^{(ex)} (E)}{N_{ch}}.
\end{equation}

Furthermore, it is demonstrated that $\rho^{(ex)}(E)$ can be written as
\begin{equation}
\rho^{(ex)}(E) = \rho^{(ex)}_0(E) + \rho^{(ex)}_{nr}(E) + \rho^{(ex)}_{res}(E),
\label{rhofin}
\end{equation}
where $\rho^{(ex)}_0(E)$, $\rho^{(ex)}_{nr}(E)$, and $\rho^{(ex)}_{res}(E)$ represent the single-channel contribution, the multichannel non-resonance term, and the multichannel resonance term, respectively~\cite{nemoto}. 
The physical meanings of these terms are clarified by introducing a ponderomotive interaction $U_{\mu\nu}(z)$ defined as 
\begin{equation}
U_{\mu\nu}(z) = \left[ F_0 z + \mu \omega \right] \delta_{\mu\nu}+V_{\mu\nu}(z) + v_{\mu\nu}(z),
\end{equation}
which includes the effect of ac-ZT. For the convenience of discussion, $U_{\mu\nu}(z)$ is classified into the single-channel part 
\begin{equation}
\bar{U}_{\mu}(z) \equiv U_{\mu\mu}(z) = F_0 z + \mu \omega +V_{\mu\mu}(z) + v_{\mu\mu}(z)
\end{equation}
termed the ponderomotive potential~\cite{nemoto}, and the interchannel one
\begin{equation}
\bar{V}_{\mu\nu}(z) \equiv (1-\delta_{\mu\nu})U_{\mu\nu}(z)  
\end{equation}
termed the ponderomotive coupling~\cite{kukuu}. 

For small $\alpha$ (i.e. for small $F_{ac}$), $\bar{U}_{\mu}(z)$ is dominant, because $\bar{V}_{\mu\nu}(z)$ is not dominant. 
Thus $\rho^{(ex)}(E)$ is dominated by $\rho^{(ex)}_0(E)$ which shows peak spectra of shape-resonance levels supported by $\bar{U}_{\mu}(z)$~\cite{karasawa}.
In particular, for $\alpha \approx 0$
\begin{equation}
 \bar{U}_{\mu}(z) \approx V(z) +F_0z + \mu\omega,
\end{equation}
and $\bar{V}_{\mu\nu}(z) \approx 0$. Then Eq.~(\ref{L}) is reduced to a decoupled equation for each $\nu$ 
and $\rho^{(ex)}(E)\approx\rho^{(ex)}_0(E)$ shows peaks corresponding to resonance states of the WSL.
As $\alpha$ increases, the shape of $\bar{U}_\mu(z)$ is more deformed and $\bar{V}_{\mu\nu}(z)$ gradually increases~\cite{kukuu}.
However the peak structure of $\rho^{(ex)}_0(E)$ is still kept unaltered to some extent.

For larger $\alpha$, $\bar{V}_{\mu\nu}(z)$ further develops and accordingly,   the contributions from the interchannel couplings,  i.e., $\rho^{(ex)}_{nr}(E)$ and $\rho^{(ex)}_{res}(E)$ are expected to become dominant.  Particularly, $\rho^{(ex)}(E)$ is governed by $\rho^{(ex)}_{res}(E)$ for large $\bar{V}_{\mu\nu}(z)$, and it is shown that $\rho^{(ex)}_{res}(E)$ can be approximately expressed as
\begin{align}
\rho^{(ex)}_{res}(E) = \sum_{\beta \tilde{\beta_i}} \frac{B_{\beta \beta_i}\bigl(E - \mathcal{E}_{\tilde{\beta}_i}\bigr) + A_{\beta \beta_i}\bigl(\Gamma_{\tilde{\beta}_i}/2\bigr)}{\bigl(E - \mathcal{E}_{\tilde{\beta}_i}\bigr)^2 + \bigl(\Gamma_{\tilde{\beta}_i}/2\bigr)^2}
\label{rhores}
\end{align}
with real values $A_{\beta \beta_i}$ and $B_{\beta \beta_i}$. 
The notation of $\tilde{\beta}_i$ is introduced to represent a set of the indexes $(\beta_i , E_{\beta_i})$, where $E_{\beta_i}$ represents the quasienergy level associated with the $\beta_i$-th single channel equation~\cite{nemoto}. 
As shown later in detail, according to Eq.~(\ref{rhores}), in the region of large $\alpha$, there is a possibility that $\rho^{(ex)}(E)$ possesses novel spectral peaks which are not attributed to those of $\rho^{(ex)}_0(E)$, where the peak is located at $E=\mathcal{E}_{\tilde{\beta}_i}$ with the resonance width $\Gamma_{\tilde{\beta}_i}/2$.
It is seen that the sum of $\rho^{(ex)}_{res}(E)$ and $\rho^{(ex)}_{nr}(E)$ is akin to the spectral function of the Shore profile~\cite{shore} with $\rho^{(ex)}_{nr}(E)$ as a background; this is an absorption lineshape of Feshbach resonance~\cite{Feshbach} characteristic of a multichannel scattering problem.
Strictly speaking, the resonance mechanism described by $\rho_{nr}^{(ex)}(E)$ and $\rho_{res}^{(ex)}(E)$ differs from the original Feshbach resonance mechanism in that in the later, a pure discrete state decays into unstructured continuum states, whereas in the former, a shape-resonance state with relatively narrow width decays into that with structured continuum states with large width~\cite{kukuu}.
Actually, even the lowest level in WSL is not completely bounded owing to a tilted potential. For this reason, the resonance governed by $\rho_{nr}^{(ex)}(E)$ and $\rho_{res}^{(ex)}(E)$ is termed "Feshbach-like" resonance in this study. 

\begin{figure}[tb]
\begin{center}
\includegraphics[width=7.cm,clip]{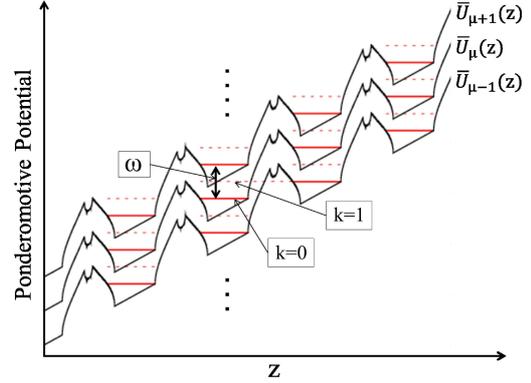}
\caption{(Color online)  Schematic representation of the multichannel-scattering character of a DFSL for $\eta=3/2$. The black solid lines represent the ponderomotive potentials $U_{\mu}(z)$ as functions of $z$. The lowest quasienergy level ($b=1$) is indicated by the red solid lines. The red dotted lines indicate the split subbands. For $\eta=3/2$, there is one split-subband in the Brillouin zone of length $\omega$.
}
\label{figlabel}
\end{center}
\end{figure}

Finally, we give a comment on the periodicity of $\rho^{(ex)}(E)$ in DFSL for $\eta=p/q$;
this quantity ensures the following periodicity relation in the Brillouin zone with an interval $\omega$ as
\begin{equation}
\rho^{(ex)}(E)=\rho^{(ex)}\left(E+n\omega+\frac{k}{q}\omega\right)
\label{rhoperiod}
\end{equation}
with $n$ and $k$ as integers, where $k$ is termed as a split-subband index in this paper with $k = 0, ..., q-1$~\cite{zhao,je}.
Fig.~\ref{figlabel} for $\eta=3/2$ clearly shows the periodicity of concern.
Let us assume that a certain potential, namely, the $\mu$-th potential $\bar{U}_\mu(z)$ associated with channel $\mu$ supports the lowest quasienergy level ($b=1$).
Then, each potential of different channel also supports a photon-sideband at the same well,
which results in the periodicity $\rho^{(ex)}(E)=\rho^{(ex)}(E+n\omega)$.
In addition, it can be found that neighboring quantum wells also support the same lowest level,
whose quasienergy levels shift by $\omega/2$, causing the periodicity $\rho^{(ex)}(E)=\rho^{(ex)}(E+\omega/2)$ for $\eta=3/2$.
Both of the periodicity are combined into the Eq.~(\ref{rhoperiod}) with $q=2$.
%, there are $q-1$ split subbands~\cite{zak} showing the same behavior as a parent mini-band labeled as $k=0$ for $\eta=p/q$. 

\section{Results and Discussion}
\label{sect3}

In this section, we present numerical results of $\rho^{(ex)}(E)$, calculated by using the R-matrix Floquet theory.
The numerical calculations are carried out for a 35/11ML-GaAs/Ga$_{0.75}$Al$_{0.25}$As SLs (1 ML = 2.83 \AA ) with a lattice constant $d=246$, where the barrier height of a confining quantum-well potential is $V_b=7.8 \times 10^{-3}$. The effective masses of electrons in the wells and the barriers are  $m_{w}=0.0665$ and $m_{b}=0.0772$, respectively~\cite{kukuu}. 
The strength of the applied bias field is set to  $F_0=104.5$ kV/cm which corresponds to the Bloch frequency of $\Omega_B = F_0 d = 5\times 10^{-3}$.  The number of quantum wells is set to $N_{qw}=20$.
All the results shown here are calculated for $N_{ch}=41$, and we have confirmed that the results sufficiently converge with respect to $N_{ch}$.

\begin{figure}[h]
\begin{center}
\includegraphics[width=5.3cm,clip]{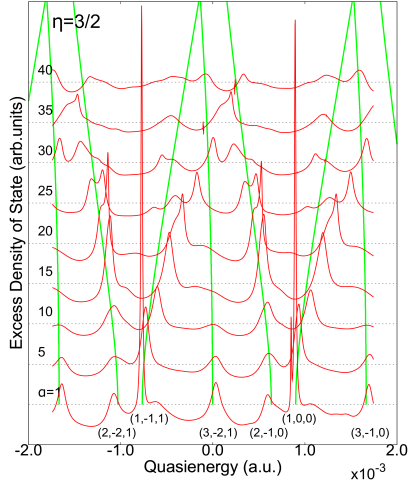}
\caption{(Color online)  The excess DOS, $\rho^{(ex)}(E)$, as a function of quasienergy, $E$, in the extended $\alpha$ region for $\eta=3/2$, which are depicted by red lines. The quasienergy position of each level is also plotted with green lines (for the sake of comparison). For the meaning of the label $(b, \mu, k)$, consult the text.
}
\label{fig3/2}
\end{center}
\end{figure}

Fig.~\ref{fig3/2} shows the calculated results of $\rho^{(ex)}(E)$ for $\eta=p/q=3/2$ from $\alpha=1$ to $40$.
We plot $\rho^{(ex)}(E)$ in one Brillouin zone $\omega=\Omega_B/\eta = 3.33 \cdots \times 10^{-3}$.
In this case, because of the periodicity of $\rho^{(ex)}(E)=\rho^{(ex)}(E+ \omega/2)$ given in Eq. (\ref{rhoperiod}), the same structure appears with the interval of $\omega/2=1.66 \cdots \times 10^{-3}$.
In the actual calculation, there are tiny differences, due to numerical inaccuracy attributed to the finite size of $N_{qw}$.

Here, each discernible peak at $\alpha=1$ is labeled as $(b,\mu,k)$. The first index $b$ labels the SL-miniband. In the present condition, the energetically lowest three states $b = 1$, $2$, and $3$ are discernible. The second one $\mu$ is the photon sideband index, 
and the last index $k$ is the split-subband index seen in Eq.(\ref{rhoperiod}).
We also plot quasienergy positions of the lowest three levels ($b=1$, $2$ and $3$) calculated by the tight binding model with intra-site ac-ZT, however, without the inter-site hoppings~\cite{TBmodel}.
The calculated quasienergy levels using the tight binding model reproduce peak positions of $\rho^{(ex)}(E)$ to some extent.
In Fig.~\ref{fig3/2}, it is found that $\rho^{(ex)}(E)$ for $\eta=3/2$ has strong $\alpha$-dependence. 
First, most of the apparent peaks at $\alpha=1$ blur with increasing $\alpha$. Some of them seem to sharpen again,
and then split around $\alpha=20$. 
The typical examples are the peaks originated from the $b=2$ miniband. 
For larger $\alpha$, those split peaks become more distant from each other. 
At $\alpha=40$, obtained peaks no longer correspond to those at $\alpha=1$.

\begin{figure}[h]
\begin{center}
\includegraphics[width=5.3cm,clip]{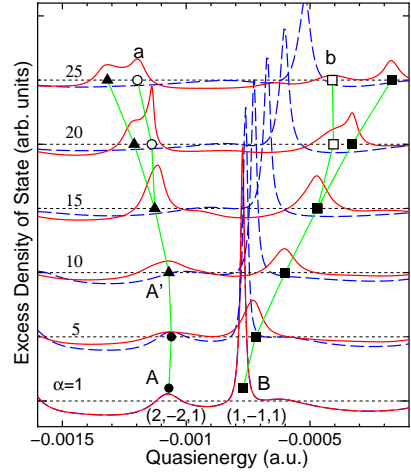}
\caption{(Color online)  
The extended figure of $\rho^{(ex)}(E)$ for $\eta=3/2$ near the peaks (2,-2,1) and (1,-1,1) from $\alpha=1$ to $25$ (red lines). The discernible peaks are connected by green lines to aid the presentation. In addition, calculated results of $\rho_0^{(ex)}(E)$ for $\eta=3/2$ are also plotted (blue dotted lines). The peaks of A, $\mr{A}^\prime$, B, a, and b are represented by a filled circle, a filled triangle, a filled square, an open circle, and an open square, respectively.
}
\label{fig3/2f}
\end{center}
\end{figure}

In Fig.~\ref{fig3/2f}, we plot $\rho^{(ex)}(E)$ near the peaks (2,-2,1) and (1,-1,1) from $\alpha=1$ to $25$ to examine $\alpha$-dependence of the peaks in more detail.
At $\alpha = 5$, the peak (2,-2,1) (labeled "A") becomes blurred. 
However, for $\alpha = 10 \sim 15$, it is seen that the intense peak (labeled "$\mr{A}^\prime$") is redshifted. 
As regards the peak (1,-1,1) labeled "B", it is found that the peak shifts to high-energy side with the increase in $\alpha$.
For $\alpha > 15$, both peaks "$\mr{A}^\prime$" and "B" show splitting accompanying new peaks labeled "a" and "b", respectively.
It is notable that the new peak "a" abruptly  develops for $15 < \alpha \le 20$, and then it is rapidly blurred for larger $\alpha$.
At $\alpha=25$, the peaks "$\mr{A}^\prime$" and "a" have similar order of peak width and height.
As regards the other new peak "b", by contrast, its width still remains broadened after the splitting from "B".
In addition, it is found that the red(blue)shift of peaks "A" and "$\mr{A}^\prime$" (peak "B") is caused by the interchannel coupling partially included in the tight binding model.

The behavior of the peaks seen in Fig.~\ref{fig3/2f} can be analyzed on the basis of the analytical expression of $\rho^{(ex)}(E)$.
As explained in the last section, the spectral peak attributed to the ponderomotive coupling $\bar{V}_{\mu\nu}(z)$ is expressed as the Feshbach-like resonance profile composed of $\rho^{(ex)}_{nr}(E)$ and $\rho^{(ex)}_{res}(E)$.
On the other hand, the rest, i.e., $\rho^{(ex)}_0(E)$, corresponds to the shape-resonance profile resulting from the ponderomotive potential $\bar{U}_{\mu}(z)$.
For the purpose of comparison, $\rho^{(ex)}_0(E)$ is also calculated by the R-matrix Floquet theory with $\bar{V}_{\mu\nu}(z)=0$, plotted by blue dashed lines in Fig.~\ref{fig3/2f}.
We can see that the broadening of the peak "A" from $\alpha=1$ to $5$ and the slight blueshift of the peak "B" are qualitatively consistent with tendency of the calculated results of $\rho^{(ex)}_0(E)$.
By contrast, in the region of $10 \le \alpha < 20$, the development of the peak "$\mr{A}^\prime$" and its redshift are not consistent with $\rho^{(ex)}_0(E)$. 
Therefore, the peak "$\mr{A}^\prime$" is considered attributed to $\rho^{(ex)}_{nr}(E)$ and $\rho^{(ex)}_{res}(E)$; for this reason, it is convenient that this label "$\mr{A}^\prime$" is distinguished from "A" attributed to $\rho^{(ex)}_0(E)$.

It would be certain that the novel peaks  "a" and "b" for $20 \leq \alpha \leq 25$ are also due to the Feshbach-like resonance mechanism.
Now, the origin of such new-peak appearance is brought to light based on the total Green function projected onto a closed channel, denoted as $G(E)$; this is associated with the denominator of the Shore profile of Eq.(\ref{rhores})~\cite{shore}. 
The Green function is formally expressed as $G(E) = 1/(E - E_0 - \Sigma(E))$,
where $E_0$ and $\Sigma(E)$ represent an unperturbed energy and a self-energy at quasienergy $E$, respectively.
It is noted that both energy and width of the Feshbach-like resonance are determined by poles of $G(E)$~\cite{shore} in the lower-half complex $E$-plane.
That is, the real part of a pole of $G(E)$ gives the resonance energy and the imaginary part gives the resonance width.
In the present system, $\Sigma(E)$ includes the effect of the ponderomotive coupling $\bar{V}_{\mu\nu}(z)$~\cite{nemoto} in a non-perturbative manner.
When $\bar{V}_{\mu\nu}(z)$ is small, $\Sigma(E)$ can be treated perturbatively, and thus $G(E)$ has a single pole at $E \approx E_0 + \Sigma(E_0)$.
However, when $\bar{V}_{\mu\nu}(z)$ is strong enough, a non-perturbative treatment is necessary,
and it is required to solve the transcendental equation of $E = E_0 + \Sigma(E)$ to determine resonance poles; the solutions generally correspond to multiple poles of $G(E)$. 
Such a scenario suggests that a Feshbach-like resonance peak originated from $\rho_{res}^{(ex)}(E)$ would split into a larger number of peaks with the increase in $\alpha$.
In particular, the peak splitting of "$\mr{A}^\prime$" and "a" can be interpreted that a single pole of $G(E)$ in the region of $\alpha \le 15$ bifurcates in the region of $\alpha \ge 20$. 

Next, a stress is put on that the peak "a" at $\alpha=20$ has extremely small $\Gamma_{\tilde{\beta}_i}$.
This seems to manifest itself under a particular condition that one of the two poles exists in the vicinity of a real axis in the lower-half complex $E$-plane. 
In general, it is known that spectral peaks are more broaden as the strength of an applied laser increases because of a power-broadening mechanism due to a strong non-linear optical effect~\cite{Murray}. 
Therefore, the irregularity observed here in spectral peak should be distinct from such a conventional understanding. 
According to our calculation, such anomalous behavior is ensured just in the development of peak "a".

\begin{figure}[h]
\begin{center}
\includegraphics[width=5.5cm,clip]{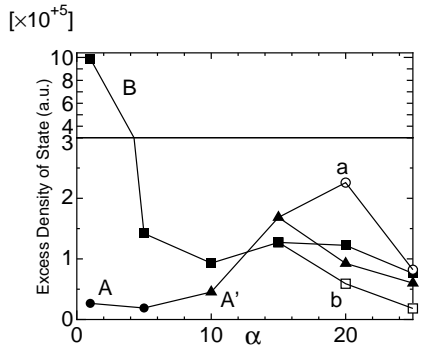}
\caption{The height of the peaks labeled in Fig.~\ref{fig3/2f}.
% In the inset, red dots are results of $\rho^{(ex)}(E)$ of $\eta=3/2$ for $\alpha=20$
% and black lines are 
}
\label{hpeaks}
\end{center}
\end{figure}

For the purpose of more quantitative discussion, the height of the labeled peaks "A", "$\mr{A}^\prime$", "B", "a", and "b" are plotted in Fig.~\ref{hpeaks}.
It is seen that the peak "$\mr{A}^\prime$" develop for $\alpha \ge 10$ and reaches the maximum with the peak height of $1.7\times 10^{5}$ at $\alpha=15$.
For $\alpha > 15$, the peak "a" develops prominently with the highest peak at $\alpha=20$.
The maxima of the peak height of "$\mr{A}^\prime$" and "a" reach about $1.7\times 10^{5}$ and $2.2\times 10^{5}$, respectively; that is, the lifetime of the peak "$\mr{A}^\prime$" is 0.10 ps and that of "a" is 0.13 ps.
These values are much larger than that of typical resonance peaks for $\eta=1$~\cite{peake}, where a ponderomotive coupling is expected to be stronger than that in the present case of $\eta=3/2$.
In this sense, we consider the manifestation of the new stable Floquet state obtained here as anomalous effect.

\begin{figure}[h]
\begin{center}
\includegraphics[width=5.5cm,clip]{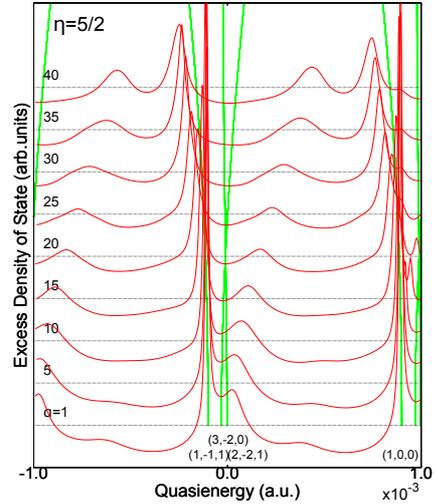}
\caption{(Color online)  The excess DOS $\rho^{(ex)}(E)$ for $5/2$
}
\label{edos_5over2}
\end{center}
\end{figure}

We also calculated $\rho^{(ex)}(E)$ for larger $\eta=5/3$ and $5/2$.
However, such novel peaks attributed to the Feshbach-like resonance are not found.
Here we plot in Fig.~\ref{edos_5over2} the results of $\rho^{(ex)}(E)$ for $\eta=5/2$; though those for $\eta=5/3$ are not shown here.
It is found that resonance peaks show just monotonic behavior in both energy-shift and width in contrast to those for $\eta=3/2$; this is consistent with results of DWSL with larger integers of $\eta>1$.
This is because the effect of $\bar{V}_{\mu\nu}(z)$ becomes negligibly smaller as $\eta$ increases~\cite{nemoto}, and thus the contribution from $\rho^{(ex)}_{res}(E)$ is less dominant.

\section{Conclusion}
\label{sect4}

In this study, we examine the resonance structure of DFSL in terms of excess DOS, $\rho^{(ex)}(E)$, by employing the R-matrix Floquet theory. 
The behavior of the obtained results of $\rho^{(ex)}(E)$ is discussed by use of the analytical expression, where this is classified into the single-channel contribution $\rho^{(ex)}_0(E)$, the multichannel non-resonance term $\rho^{(ex)}_{nr}(E)$, and the multichannel resonance term $\rho^{(ex)}_{res}(E)$. 

It is found that the ac-ZT included in the ponderomotive couplings plays an important role of the spectral property of $\rho^{(ex)}(E)$ for $\eta=3/2$, where an irregular pattern of it, especially, an anomalous Floquet state with longevity manifests itself.
The underlying physics of this manifestation is discussed in the light of the Feshbach-like resonance profile described by both $\rho^{(ex)}_{nr}(E)$ and $\rho^{(ex)}_{res}(E)$, and poles of the associated Green function $G(E)$.
In particular, the peak splitting of "$\mr{A}^\prime$" and "a" at $\alpha=20$ seems to result from the bifurcation of pole of $G(E)$ due to the non-perturbative effect of $\bar{V}_{\mu\nu}(z)$.
Further, the origin of the narrow width of the peak "a" at $\alpha=20$ is also discussed in terms of $G(E)$.
On the other hand, long-lived states are generally extremely rare in such a high laser-intensity region due to the effect of power-broadening.
Actually, such anomalous peaks are not observed for other $\eta$'s to the best of one's knowledge. 

Finally, we give a comment on a possibility of experimental observation of the finding obtained here. 
The laser intensity at $\alpha=20$ for $\eta=3/2$ is $F_{ac}=88$ kV/cm, where the prominent peak of "a" is discerned.
This is feasible strength of the terahertz cw laser in experiments at present~\cite{Hirori}, and thus it apparently seems that the corresponding peak might be observed in semiconductor SLs.
However, in actual experiments, the peak would be smeared by various sorts of broadening mechanisms due to impurities, many-body interactions, lattice imperfection, lattice vibration, and so on. 
Therefore, a more ideal system is preferred.
One possible candidate is an optical SLs composed of cold atomic gas~\cite{gluck,Madison1,Madison2}, where these perturbations would be suppressed to a great extent, and thus, the quantum phenomenon of concern is expected to be realized.

%In addition, we confirmed that the ponderomotive coupling effect becomes irrelevant as the number of $\eta$ increases by examining $\rho^{(ex)}(E)$ for $5/2$. 

% estimation by nemoto
%/eta=3/2 \omega=90[meV] -> 21.76[THz]
%      \alpha=1~40 ->Fac��4.4~176 kV/cm
%
%\eta=5/2 \omega=54.4[meV] ->13.15[THz]
%      \alpha=1~40 -> Fac��1.6~63.4 kV/cm
%
%\eta==5/3 \omega=81.6[meV] -> 19.173THz]
%      \alpha=1~40 -> Fac��3.6~142.6 kV/cm
%

{\bf Acknowledgments}

\noindent

This work was supported by Grants-in-Aids for Scientific Research on Innovative Areas ``Optical science of dynamically correlated electrons (DYCE)'' (Grant No. 21104504)
of the Ministry of Education, Culture, Sports, Science and Technology (MEXT), Japan.


\appendix
\section{Floquet theorem}
\label{sec_app_A}

In this appendix, we provide a brief introduction of the Floquet theorem~\cite{grifoni}.
Let us consider a system whose Hamiltonian ${\mathcal H}(z,t)$ has the time-periodicity
\begin{equation}
  {\mathcal H}(z,t+T) = {\cal H}(z,t).
\end{equation}
According to the Floquet theorem, there is a solution of the Sch\"{o}dinger equation~(\ref{barH}) in the form
\begin{equation}
  \Psi(z,t) = e^{-iEt} \Psi'(z,t),
  \label{eq_Floa}
\end{equation}
where $\Psi'(z,t)$ is the Floquet wavefunction, which satisfies the periodicity
\begin{equation}
  \Psi'(z,t+T) = \Psi'(z,t),
\end{equation}
and $E$ is the quasienergy. Then we substitute a Fourier expansion of the time-periodic function
\begin{equation}
 \Psi'(z,t) = \sum_{\nu =- \infty}^{\infty} \exp{(i\nu \omega t)}\psi_\nu(z)
\end{equation}
into Eq.~(\ref{eq_Floa}) and obtain Eq.~(\ref{eq_fourier}).
In actual calculation, the summation of the index $\nu$ is limited in the finite range from $-N_{ph}$ to $N_{ph}$.

%%%%%%%%%%%%%%%%%%%%%%%%%%%%%%%%%%%%%%%%%%%%%%%%%%%%%%%%%%%%%%%%%
\section*{References}


\begin{thebibliography}{00}
\bibitem{ochan} Exactly speaking, $N_{ch}$ in this paper must be replaced by $N_o$, the number of open channels.
Here a open channel labeled by $\nu$ is a channel with $(E- \nu \omega - F_0z - V_{as})>0$ in the asymptotic region. 
Then the solution of the asymptotic equation~(\ref{eq_aseq}) is a progressive wave.
In actual calculations, the number of channels $N_{ch}=2N_{ph}+1$ is finite, and thus all the channels can become the open channels
in the asymptotic region of $|z|>>L_{SL}$ with negative $z$. So we can set $N_{ch}=N_o$.  For more detail, consult Refs.~\cite{exdos1,exdos2,exdos3}.
\bibitem{TBmodel} The tight-binding model used here is basically same as that of Ref.~\cite{nemoto}. The difference is that the nearest-neighbor inter-site (well) hoppings are neglected in this work.
For the DFSL, the coupling is negligibly small because the photon-assisted tunneling between nearest wells does not occurs since $\omega\ne \Omega_B$
\end{thebibliography}


\end{document}